\documentclass{article}
\usepackage{wasysym} %needed for \hexstar
\usepackage{pifont}
\usepackage{epsfig}
\usepackage{citesort}
\usepackage{jmb}
\usepackage{citesupernumber}

\begin{document}

\title{Multiple Folding Pathways of the SH3 domain}
\author{Jose M. Borreguero$^1$\thanks{Corresponding author.
Email: jmborr@bu.edu},
Feng Ding$^1$, Sergey V. Buldyrev$^1$, H. Eugene Stanley$^1$, \\ 
and Nikolay V. Dokholyan$^{2}$}

\maketitle

\begin{center}
{\it {\small
$^1$Center for Polymer Studies and Department of Physics, 

Boston University, Boston, MA 02215 USA 

$^2$ Department of Biochemistry and Biophysics, School of Medicine,

University of North Carolina at Chapel Hill, Chapel Hill, NC 27599 
}}
\end{center}

\begin{abstract} 
  Experimental observations suggest that proteins follow different pathways
  under different environmental conditions. We perform molecular dynamics
  simulations of a model of the SH3 domain over a broad range of
  temperatures, and identify distinct pathways in the folding transition.  We
  determine the kinetic partition temperature ---the temperature for which
  the SH3 domain undergoes a rapid folding transition with minimal kinetic
  barriers--- and observe that below this temperature the model protein may
  undergo a folding transition via multiple folding pathways. The folding
  kinetics is characterized by slow and fast pathways and the presence of
  only one or two intermediates. Our findings suggest the hypothesis that the
  SH3 domain, a protein for which only two-state folding kinetics was
  observed in previous experiments, may exhibit intermediates states under
  extreme experimental conditions, such as very low temperatures. A very
  recent report (Viguera et al., Proc.  Natl.  Acad.  Sci. USA,
  100:5730--5735, 2003) of an intermediate in the folding transition of the
  Bergerac mutant of the $\alpha$-spectrin SH3 domain protein supports this
  hypothesis.\\
\end{abstract}

{\it Keywords}: intermediate, molecular dynamics, folding pathways, SH3,
kinetic partition.

\twocolumn

\section{INTRODUCTION}
  
Recent experimental studies indicate that several proteins exhibit
simultaneously a variety of intermediates and folding pathways.
Kiefhaber\cite{Kiefhaber95} identified at low denaturant concentration a
fast pathway (50 ms) in the folding of lyzosyme with no intermediates and a
slow phase (420 ms) with well-populated intermediates.  Choe et
al.\cite{Choe98} observed the formation of a kinetic intermediate in the
folding of villin 14T upon decreasing the temperature, and Silverman et
al.\cite{Silverman00} observed the extinction of a slow phase in the folding
of the P4-P6 domain upon changes in ion concentration.  Kitahara et
al.\cite{Kitahara03} studied a pressure-stabilized intermediate of ubiquitin,
identified as an off-pathway intermediate in previous kinetics experiments at
basic conditions\cite{Briggs92}. All these studies suggest that environmental
conditions favor some folding pathways over others.  

Major theoretical efforts in the study of protein
folding\cite{Borreguero02,Ding02,Fersht02,Karplus02,Ozkan02,Plotkin02,Thirumalai02,Tiana01,Pande00,Eaton00,Bryngelson89}
have focused on small, single domain proteins~\cite{Jackson98}. It is found
in experiments\cite{Jackson98,Filimonov99} that these proteins undergo
folding transition with no accumulation of kinetic intermediates in the
accessible range of experimental conditions.  However, other kinetics studies
of two-state proteins\cite{Chiti99,Houliston02,Khorasanizadeh96,Bachmann01}
suggest the presence of short-lived intermediates that cannot be directly
detected experimentally. Recently, S\'anchez et al.\cite{Sanchez03} explained
the curved Chevron plots --- the nonlinear dependence of folding and
unfolding rates on denaturant
concentration\cite{Ikai73,Matouschek90,Fersht00b} --- of 17 selected
proteins by assuming the presence of an intermediate state. Led by these
studies, we hypothesize that single domain proteins may exhibit intermediates
in the folding transition under suitable environmental conditions.

To test our hypothesis, we perform a molecular dynamics study of the folding
pathways of the c-Crk SH3 domain\cite{Wu95,Branden99,Berman00}
(PDB\cite{Berman00} access code 1cka).  The SH3 domain is a family of small
globular proteins which has been extensively studied in kinetics and
thermodynamics
experiments\cite{Viguera94,Villegas95,Grantcharova97,Grantcharova98,Knapp98,Martinez99,Filimonov99,Riddle99,Guerois00}.
We select the c-Crk SH3 domain ($57$ residues) as the SH3 domain
representative and perform molecular dynamics simulations over a broad range
of temperatures. We determine the {\it kinetic partition
  temperature}\cite{Klimov96,Thirumalai02} $T_{KP}$ below which the model
protein exhibits slow folding pathways and above which the protein undergoes
a cooperative folding transition with no accumulation of intermediates. Below
$T_{KP}$, we study the presence of intermediates in the slow folding pathways
and resolve their structure. We find that one of the intermediates populates
the folding transition for temperatures as high as $T_{KP}$. We discuss
  the relevance of our results in light of recent experimental evidence.

\section{RESULTS}

The SH3 domain is a $\beta$-sheet protein (upper triangle of
Fig.~\ref{fig:fig1}a). Our previous thermodynamic studies\cite{Borreguero02}
of the c--Crk SH3 domain revealed only two stable states at equilibrium
conditions: folded and unfolded. Both states coexist with equal probability
at the folding transition temperature, $T_F=0.626$, at which the temperature
dependence of the potential energy has a sharp change, and the specific heat
has a maximum (experimentally\cite{Filimonov99}, this temperature corresponds
to $67^\circ$C).  Thus, our model reproduces the experimentally-determined
thermodynamics of the SH3 domain\cite{Viguera94,Villegas95,Filimonov99}.

\vspace{4mm}
\subsection{Initial Unfolded Ensemble}

\begin{figure}[htb]

  \centerline{  
    \epsfig{file=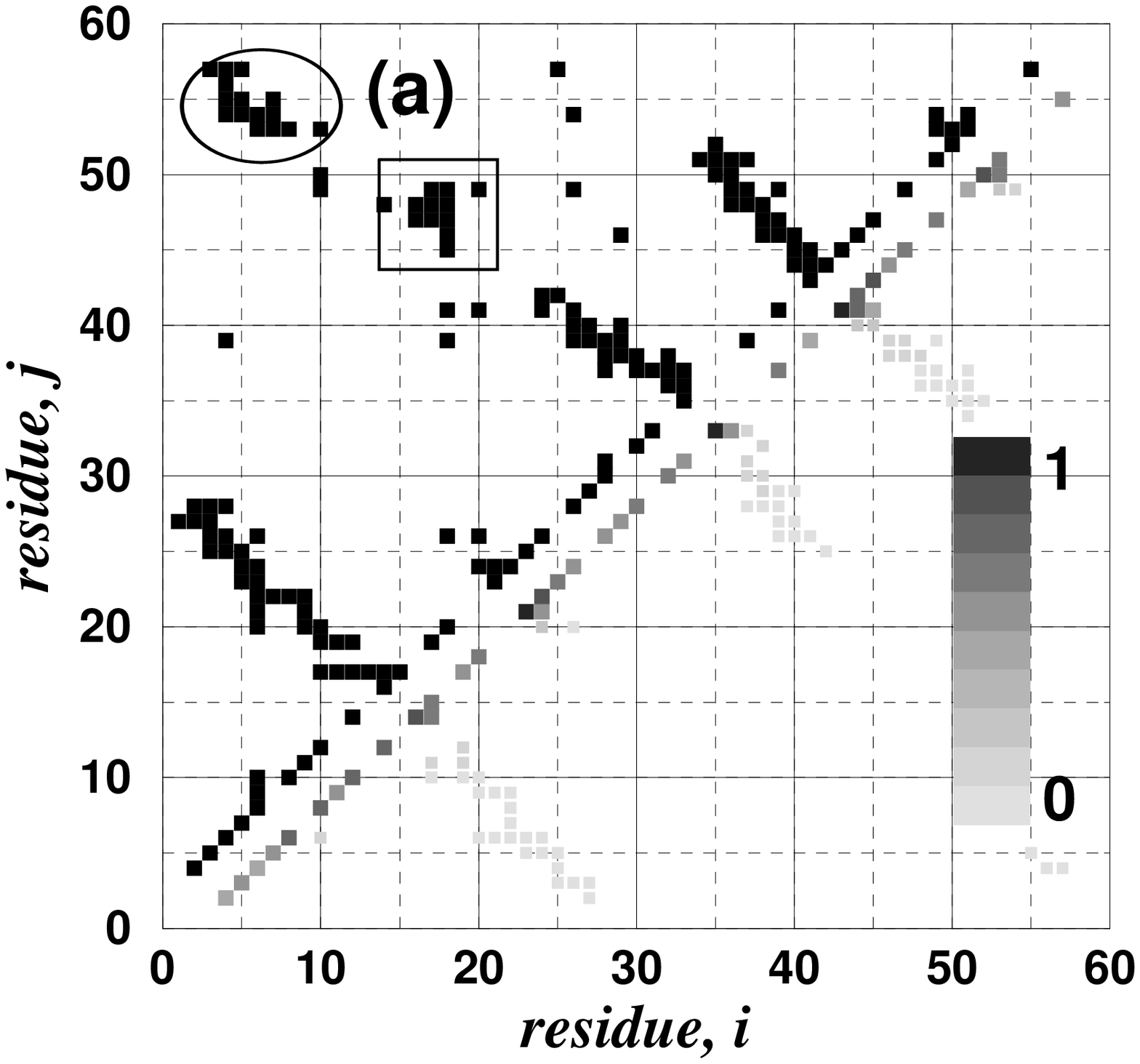, height=80mm, width = 80mm} 
    \epsfig{file=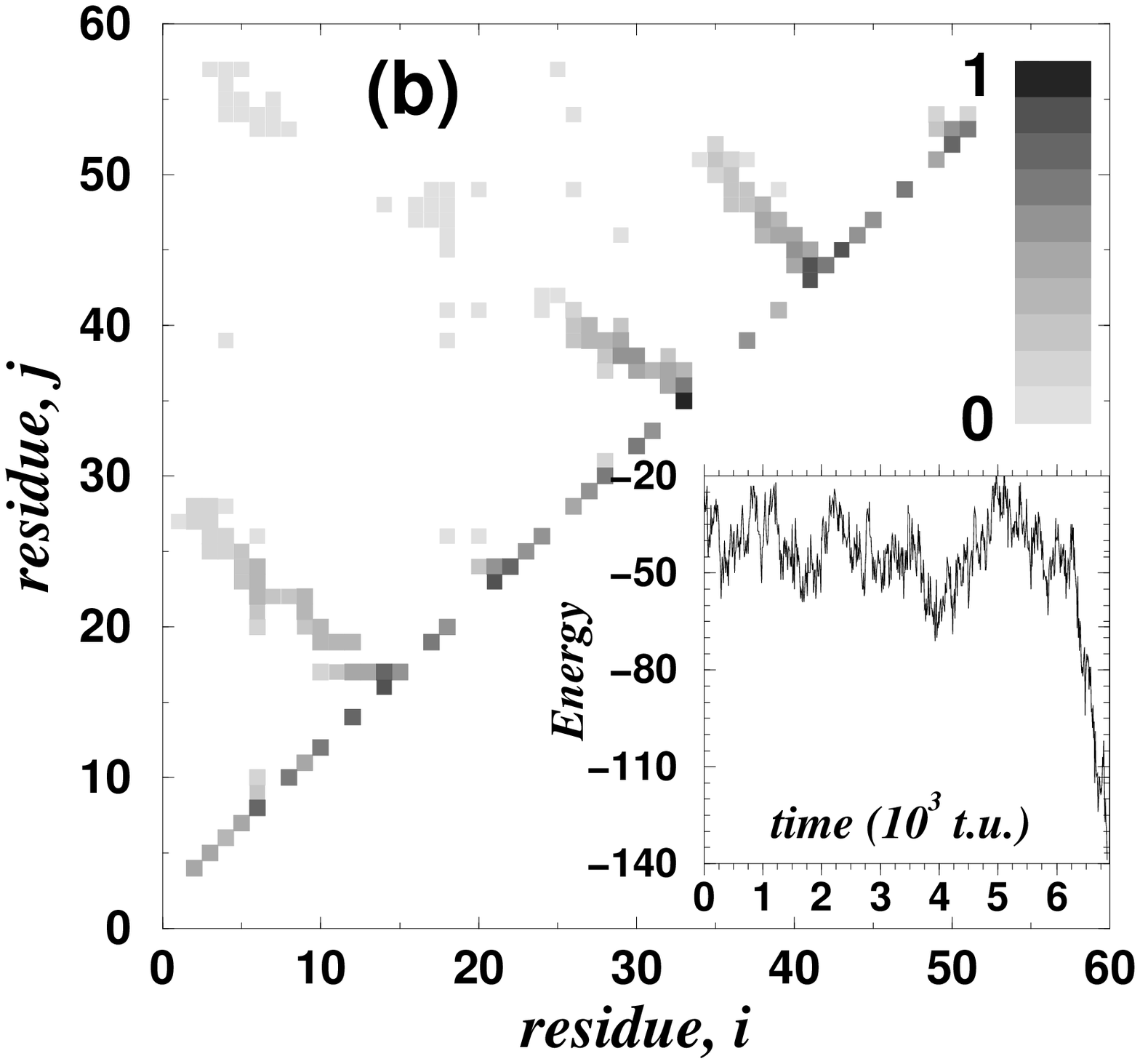, height=80mm, width = 80mm}
  }
  \centerline{
    \epsfig{file=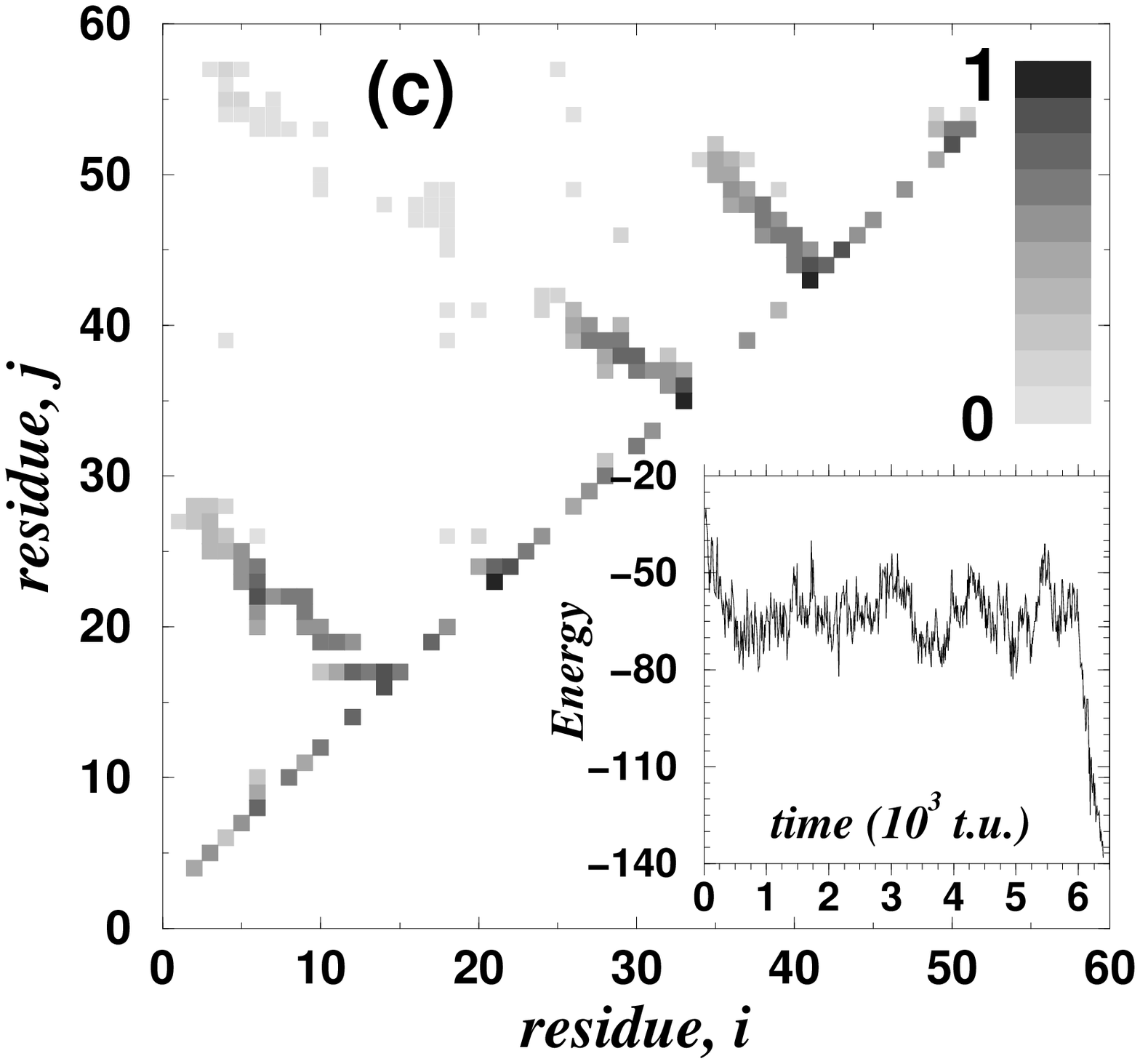, height=80mm, width = 80mm}
    \epsfig{file=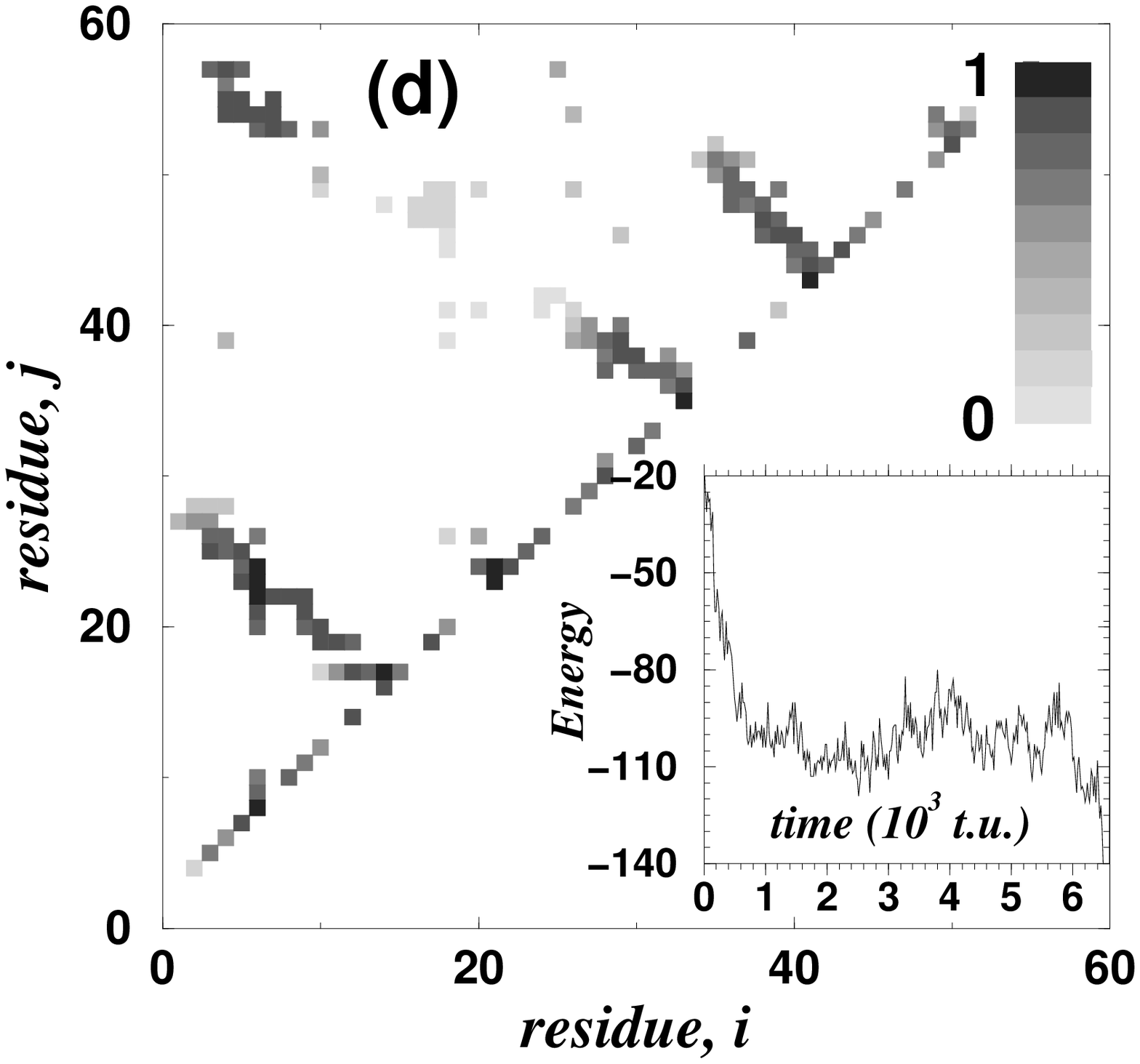, height=80mm, width = 80mm}
  }
\caption[]{
}
\label{fig:fig1}
\end{figure}

Our initially unfolded ensemble consists of 1100 protein conformations that
we sample from a long equilibrium simulation at a high temperature $T_0=1.0$
at equal time intervals of $10^4$ time units (t.u). This time separation is long
enough to ensure that the sampled conformations have low structural
similarity among themselves. We calculate the frequency map --- the plot of
the probability of any two amino acids forming a contact --- of this unfolded
ensemble (lower triangle of Fig.~\ref{fig:fig1}a). At $T=1.0$, only nearest
and next nearest contacts have high frequency, and the frequency decreases
rapidly with the sequence separation between the amino acids. 

When we quench the system from $T=1.0$ to a target temperature, $T_{\rm
  target}$ (see Methods), the system relaxes in approximately $1500$ t.u.
Due to the finite size of our heat bath, the heat released by the protein
upon folding increases the final temperature of the system by $0.03$ units
above $T_{\rm target}$. After relaxation, the protein stays for a certain
time in the unfolded state, then undergoes a folding transition.  During this
time interval, the protein explores unfolded conformations, and we calculate
the frequency map of the unfolded state for different target temperatures.

At $T_{\rm target}=T_F$, the secondary structure is unstable
(Fig.~\ref{fig:fig1}b), with average frequency $\bar{f}=0.24$ (see Methods).
Successful folding requires the cooperative formation of contacts throughout
the protein in a nucleation process\cite{Borreguero02,Ding02}. At $T_{\rm
  target}=0.54$, the secondary structure is more stable
(Fig.~\ref{fig:fig1}c, $\bar{f}=0.50$).  Thus, the conformational search for
the native state (NS) is optimized by limiting the search to the formation of
a sufficient number of long range contacts.  At $T_{\rm target}=0.33$, the
lowest temperature studied, secondary structure elements form during the
rapid collapse of the model protein in the first $1500$ t.u.
(Fig.~\ref{fig:fig1}d, $\bar{f}=0.73$). During collapse, some tertiary
contacts --- contacts between secondary elements --- may also form.  The
formation of these contacts prior to the proper arrangement of secondary
structure elements may lead the protein model to a kinetic trap.  Finally,
folding proceeds at this temperature through a thermally activated search for
the NS.

\subsection{Kinetic Partition Temperature}

\begin{figure}[htb]
  
  \centerline{ 
    \epsfig{file=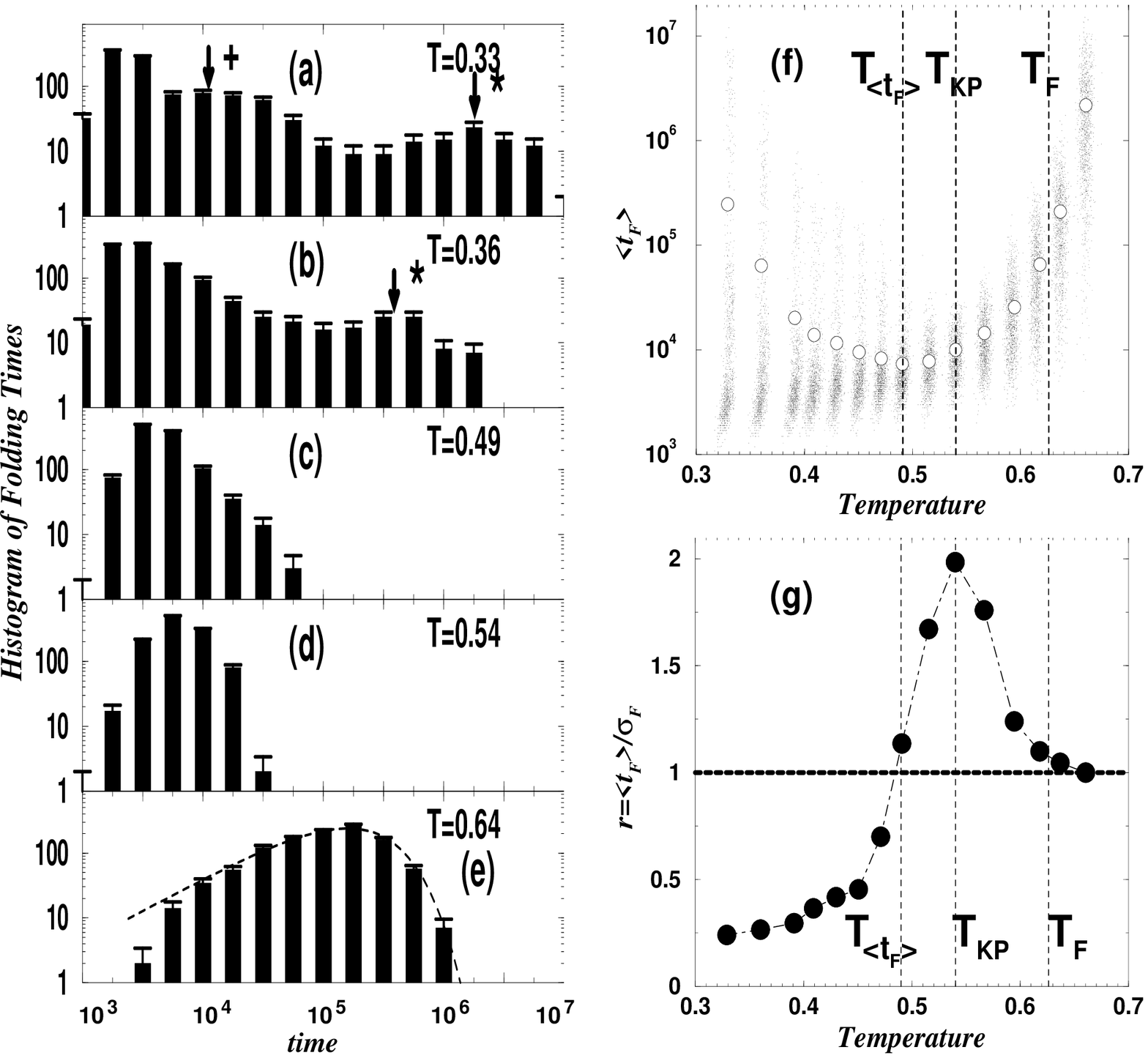, height=160mm, width=160mm}
  }
    
\caption[]{
}
\label{fig:fig2}
\end{figure}

In order to determine the temperature below which we can distinguish fast and
slow folding pathways, we compute the distribution of folding times
$p(t_F,T)$ (Fig.~\ref{fig:fig2}a--e), as well as the average $\langle t_F
\rangle$ (Fig.~\ref{fig:fig2}f) and standard deviation $\sigma_F$. The ratio
$r(T) \equiv \langle t_F \rangle/\sigma_F$ measures the average folding time
in units of the standard deviation $\sigma_F$. This quantity is particularly
useful when the value of the standard deviation correlates with the value of
the average as we change $T_{\rm target}$.  For instance, single-exponential
distributions $e^{-t_F/\langle t_F \rangle}/\langle t_F \rangle$ have $r
\equiv 1$. 

We expect $r \rightarrow 1$ for $T_{\rm target} > T_F$, because at these high
temperatures the folding transitions become rare events and are
single-exponential distributed. As we decrease $T_{\rm target}$, we expect $r
> 1$ just below $T_F$, because the folded state becomes more stable than the
unfolded state, and the folding transitions are favored. Distributions with
$r > 1$ indicate a narrow distribution centered in $\langle t_F\rangle$, so
that most of the simulations undergo a folding transition for times of the
order of the average folding time. However, if we continue decreasing $T_{\rm
  target}$, we expect some folding transitions to be kinetically trapped, and
the folding time distribution will spread over several orders of magnitudes.
Such distributions have $r < 1$. Thus, there is a temperature below $T_F$
where the maximum of $r(T)$ occurs, and which signals the onset of slow folding
pathways. We use the maximum of $r(T)$ to calculate $T_{KP}$.

Fig.~\ref{fig:fig2}g suggests that $T_{KP}=0.54$, which corresponds to a
maximally compact distribution of folding times\footnote{Assuming a linear
  relation between experimental and simulated temperatures and taking into
  account\cite{Filimonov99} that $T_F=67^\circ$C, we estimate $T_{KP} \simeq
  20^\circ$C} (Fig.~\ref{fig:fig2}d).  We find that the ratio approaches one
as we increase the temperature above $T_{KP}$, and the distribution of
folding times approximates a single-exponential distribution. In particular,
the distribution of folding times fits the single-exponential distribution
$e^{-t_F/\langle t_F \rangle}/\langle t_F \rangle$ for $T=0.64$, the closest
temperature to $T_F$ that we study.  The ratio $r(T)$ decreases
monotonically below $T_{KP}$, indicating that the distribution of folding
times spreads over several orders of magnitude. This is the consequence of an
increasing fraction of folding simulations kinetically trapped
(Fig.~\ref{fig:fig2}a-b).  The average folding time $\langle t_F \rangle$ is
minimal not at $T_{KP}$, but at a lower temperature $T_{\langle t_F
  \rangle}=0.49$ (Fig.~\ref{fig:fig2}f). At this temperature, we find that
the protein becomes temporarily trapped in approximately $7\%$ of the folding
transitions. On the other hand, the remaining simulations undergo a folding
transition much faster, thus minimizing $\langle t_F \rangle$.
Interestingly, $r(T_{\langle t_F \rangle}) \simeq 1.0$, even though the
distribution of folding times at this temperature is non-exponential.

\subsection{Folding Pathways}

Below $T_{KP}$, an increasing fraction of the simulations undergo folding
transitions that take a time up to three orders of magnitude above the
minimal $\langle t_F \rangle$. In addition, $\langle t_F \rangle$ increases
dramatically (Fig.~\ref{fig:fig2}f). At the lowest temperatures studied, we
distinguish between the majority of simulations that undergo a fast folding
transition (the fast pathway) and the rest of the simulations that undergo
folding transitions with folding times spanning three orders of magnitude
(the slow pathways). At the low temperature $T=0.33$, the potential energy of
the fast pathway has on average the same time evolution of all the
simulations at $T_{KP}=0.54$, indicating that there are no kinetic traps in
the fast pathway.

For each folding simulation that belongs to the slow pathways, we sample the
potential energy at equal time intervals of $100$ t.u. until folding is
finished (see Methods). Then, we collect all potential energy values and
construct a distribution of potential energies. We find that below $T=0.43$,
the distribution is markedly bimodal (Fig.~\ref{fig:fig3}a). The positions of
the two peaks along the energy coordinate do not correspond to the
equilibrium potential energy value of the folded state
(Fig.~\ref{fig:fig3}b). Therefore we hypothesize the existence of two
intermediates in the slow pathways. We denote the two putative intermediates
as $I_1$ and $I_2$ for the high energy and low energy peaks, respectively. As
temperature decreases, the peaks shift to lower energies, but the energy
difference between the two peaks, approximately six energy units, remains
constant (Fig.~\ref{fig:fig3}b). A constant energy difference implies that
the two putative intermediates differ by a specific set of native contacts.
As temperature decreases, other contacts not belonging to this set become
more stable and are responsible for the overall energy decrease.  At
$T=0.33$, we record the distribution of survival times for both intermediates
and find that they fit a single-exponential distribution, supporting the
hypothesis that each intermediate is a local free energy minima and has a
major free energy barrier (Fig.~\ref{fig:fig3}c).

To further test the single free energy barrier hypothesis, we select a
typical conformation representing intermediate $I_2$ and perform 200 folding
simulations, each with a different set of initial velocities for a set of
temperatures in the range $0.33 \leq T \leq 0.52$.  For each simulation, we
record the time that the protein stays in the intermediate and find that the
average survival time fits the Arrhenius law for temperatures below $T=0.44$
(Fig.~\ref{fig:fig3}d). This upper bound temperature roughly coincides with
the temperature $T=0.43$ below which $I_2$ becomes noticeable in the
histogram of potential energies (Fig.~\ref{fig:fig3}a).  This result
indicates that the free energy barrier to overcome intermediate $I_2$ becomes
independent of temperature for low temperatures, or analogously, that the
same set of native contacts must form (or break) to overcome the
intermediate.

Next, we determine the structure of the two intermediates. For each
intermediate, we randomly select three conformations and find that they are
structurally similar. Conformations belonging to intermediate $I_1$ have a
set of long-range contacts (C$_1$) with a high occupancy and a set of
long-range contacts (C$_2$) with no occupancy at all (Fig.~\ref{fig:fig3}e).
Contacts in C$_1$ represent a $\beta$-sheet made by three strands: the two
termini and the strand following the RT-loop, which we name strand ``A'' (see
$I_1$ in Fig.~\ref{fig:pathCartoon}). Contacts in C$_2$ represent the base of
the n-Src loop and the contacts between the RT-loop and the distal hairpin
(see $I_2$ in Fig.~\ref{fig:pathCartoon}). In addition, $I_1$ has a set of
medium-range contacts (C$_3$) with high occupancy (Fig.~\ref{fig:fig3}e)
representing the distal hairpin and a part of the n-Src loop. For a slow
folding transition, the $\beta$-sheet (C$_1$ contacts) forms in the early
events and strand ``A'' can no longer move freely. This constrained motion
prevents strand ``A'' from forming contacts with one of the strands of the
distal hairpin, which we name as strand ``B'' (see $I_2$ in
Fig.~\ref{fig:pathCartoon}).  Similarly, strand ``B'' cannot move freely
because it is a part of C$_3$. The missing contacts between strand ``A'' and
strand ``B'' are the contacts that form the base of the n-Src loop.

Conformational changes leading the protein away from intermediate $I_1$
involve either dissociation of the $\beta$-sheet, thus breaking some contacts
of C$_1$, or dissociation of the distal hairpin, thus breaking some contacts
of C$_3$.  We find that the latter dissociation may lead the protein
conformation to intermediate $I_2$.  Intermediate $I_2$ has contacts of
C$_1$, but lacks the set of contacts (C$_4$) that form the base of the the
distal hairpin (see NS in Fig.~\ref{fig:pathCartoon}).

Once we identify the structure of the intermediates, we investigate whether
intermediate $I_1$ is present at larger temperatures when no distinction can
be made concerning fast and slow folding pathways. To test this hypothesis,
we sample the protein conformation during the folding transition at equal
time intervals of $60$ t.u. for each of the $1100$ simulations, and compare
these conformations to intermediate $I_1$ with a similarity score function
(see Methods). For each folding transition, we record only the highest value
of the similarity score, thus obtaining $1100$ highest score values. At
$T_{KP}$, the histogram of the highest scores is bimodal, with $25\%$ of the
folding simulations passing through intermediate $I_1$
(Fig.~\ref{fig:fig3}f). We find that at $T_{KP}$, simulations that undergo
the folding transition through $I_1$ show kinetics of folding no different
than those of the rest of simulations.

\section{DISCUSSION}

It was shown\cite{Borreguero02} that the simplified protein model and
interaction potentials that we use here reproduced in a certain range of
temperatures the\\ \mbox{experimentally-determined} two-state thermodynamics
of the SH3 domain\cite{Filimonov99}. The qualitative predictive power of the
model encouraged us to study the folding kinetics in a wider range of
temperatures.  From our relaxation studies of the initial unfolded ensemble,
we observe that the structure of the unfolded state is highly sensitive to
$T_{\rm target}$.  The role of the unfolded state in determining the folding
kinetics has already been pointed out in recent experimental and theoretical
studies\cite{Zagrovic02,Millet02,Plaxco01,Garcia01}. We observe
nucleation\cite{Borreguero02}, folding with minimal kinetic barriers, and
thermally activated mechanisms for the different observed unfolded states.

In previous studies, various methods have been developed to determine the
temperature that signals the onset of slow folding pathways. Socci et
al.\cite{Socci94,Socci96} determined a {\it glass transition temperature},
$T_g$, at which the average folding time is half way between $t_{\rm min}$
and $t_{\rm max}$, where $t_{\rm min}$ is the minimun average folding time
and $t_{\rm max}$ is the total simulation time. This method is sensitive to
the {\it a priori} selected $t_{\rm max}$. The authors varied $t_{\rm max}$
in the range $0.27 \times 10^9 < t_{\rm max} < 0.960 \times 10^9$, and they
found a $10\%$ error in the calculation of $T_g$. Also, Gutin et
al.\cite{Gutin98b} estimated a {\it critical temperature}, $T_c$, at which
the temperature dependence of the equilibrium potential energy leveled off.
From their results, one can evaluate a $20\%$ error in their calculation of
$T_c$. Both $T_g$ and $T_c$ are temperatures that authors use to characterize
the onset of multiple folding pathways. In our study we use $T_{KP}$, which
signals the breaking of time translational invariance of equilibrium
measurements for temperatures below this value\cite{Dokholyan02}. We estimate
a $2\%$ error in our calculation of $T_{KP}$ from uncertainties in the
location of $T_{KP}$ in Fig.~\ref{fig:fig2}g.

At $T_{KP}$, secondary structure elements are partially stable, and the
search for the NS reduces to the formation of tertiary contacts.
Furthermore, $T_{KP}$ is a relatively high temperature that prevents the
stabilization of improper arrangements of the protein conformation, thus
minimizing the occurrence of kinetic traps. Below $T_{KP}$, the model protein
exhibits two intermediates with well-defined structural characteristics. This
modest number of intermediates is a direct consequence of the prevention of
non-native contacts. This prevention reduces dramatically the number of
protein conformations.  Furthermore, since a low energy value implies that
most of the native interactions have formed, there are few conformations
having both low energy and structural differences with the NS\cite{Plotkin02}. 

It is found experimentally
\cite{Bhutani01,Tezcan02,Juneja02,Choe98,Park97,Kiefhaber95,Silverman00,Segel99,Heidary00,Rumbley01,Simmons02}
that proteins exhibit only a discrete set of intermediates. Even though in
real proteins amino acids that do not form a native contact may still attract
each other, experimental and theoretical studies confirm that native contacts
have a leading role in the folding transition.  Protein engineering
experiments\cite{Fersht95,Grantcharova98,Ternstrom99,Clementi00,Northey02}
show that transition states in two-state globular proteins are mostly
stabilized by native interactions. To quantitatively determine the importance
of native interactions in the folding transition, Paci et al.\cite{Paci02}
studied the transition states of three two-state proteins with a full-atom
model. They found that on average, native interactions accounted for
approximately $83\%$ of the total energy of the transition states. Of
relevance to our studies of the SH3 domain are the full-atom
study\cite{Shea02} and the protein engineering
experiments\cite{Grantcharova98,Riddle99} showing that the transition state
of the src-SH3 domain protein is determined by the NS.  On the other hand,
evidence exists that in some proteins, non-native contacts are responsible
for the presence of intermediates. In their study of the homologous Im7 and
Im9 proteins, Capaldi et al.\cite{Capaldi02} identified a set of non-native
interactions responsible for a intermediate state in the folding transition
of Im7 protein. Mirny et al.\cite{Mirny96b} performed Monte Carlo simulations
of two different sequences with the same NS in the $3\times 3$ lattice.  One
sequence presented a series of pathways with misfolded states due to
non-native interactions.
  
We investigate the kinetics of formation of the two intermediates in a wide
range of temperatures. At low temperatures, simulations that undergo folding
through intermediate $I_1$ reveal that contacts between the two termini form
earlier than the contacts belonging to the folding
nucleus\cite{Borreguero02,Ding02}. This result coincides with an off-lattice
study\cite{Abkevich94b} of a 36-monomer protein by Abkevich et al. In this
study, the authors found an intermediate in the folding transition of their
model protein. Inspection of the intermediate revealed no nucleus contacts,
but a different set of long-range contacts already formed. In addition, we
learned of the work by Viguera et al.\cite{Viguera03} after completion of our
study. They reported that a mutant of the $\alpha$-spectrin SH3 domain
undergoes a folding transition through one intermediate. The authors observed
that the newly-introduced long-range contacts had already been formed in the
denaturated state, preceding the formation of the transition state of the
protein. Thus, environmental conditions that favor stabilization of long-range
contacts other than the nucleus contacts may induce intermediates in the
folding transition. 

Alternatively, short-range contacts in key positions of the protein structure
may also be responsible for slow folding pathways. After completion of our
study, Karanicolas et al.\cite{Karanicolas03,Nguyen03} reported their studies
on the G\={o} model of the forming binding protein WW domain. The authors
found a slow folding pathway in the model protein, and a cluster of four
short-range native contacts that are responsible for this pathway. However,
the authors observed that it was the absence, not the presence, of these
native contacts in the unfolded state that generated bi-phasic folding
kinetics. Thus, environmental conditions that favor destabilization of short
range contacts may promote the formation of intermediate states in the
folding transition.

We also investigate the survival time of intermediate $I_2$, and find that
the free energy barrier separating $I_2$ from the NS is independent of
temperature. Thus, the average survival time follows Arrhenius kinetics.  The
value of the free energy barrier is approximately $5.85$ energy units,
indicating that about six native contacts break when the protein conformation
reaches the transition state that separates $I_2$ from the NS. At the low
temperatures where intermediates $I_1$ and $I_2$ are noticeable, thermal
fluctuations are still large enough so that the observed survival times of
$I_2$ should be much smaller, if only any six native contacts were to break.
Thus we hypothesize that it is allways the same set of native contact that
must break in the transition $I_2 \rightarrow NS$. Our observations of the
transition $I_1 \rightarrow I_2$ support this hypothesis. In this transition,
we find that the set of contacts C$_4$ allways breaks.

At $T_{KP}$, we do not detect the intermediates from kinetics measurements of
the average folding time, or analogously, from the folding rate. Thus we
analyze the folding transition with the similarity score function that tests
the presence of intermediate $I_1$. Then we find that this intermediate is
populated in $25\%$ of the folding transitions. In a recent
study\cite{Gorski01}, Gorski et al.  reported the existence of an
intermediate in the folding transition of protein Im9 under acidic conditions
($pH=5.5$). This finding led authors to formulate the hypothesis that Im9 has
an intermediate at normal conditions ($pH=7.0$), but it is too unstable to be
detected with current kinetic experimental techniques. Interestingly, the
homologous protein Im7 ($60\%$ sequence identity) undergoes folding
transition through an intermediate in all tested experimental
conditions\cite{Ferguson99,Gorski01,Capaldi02}, supporting the authors'
hypothesis. Thus, changes in both the environmental conditions and the amino
acid sequence may uncover hidden intermediates in the folding transition of a
two-state protein. In addition, a detailed study at $T_{KP}$ may reveal the
intermediates.  This is particularly useful for computer simulations, because
simulations at low temperatures when intermediates are easily identifiable
may require several orders of magnitude longer than simulations at $T_{KP}$.

%\newpage
\vspace{4mm}
\subsection{Conclusion}

We perform molecular dynamics analysis of the folding transition of the
G\={o} model of the c-Crk SH3 domain in a broad range of temperatures.  At
the folding transition temperature, we observe that only the folded and
unfolded states are populated.  However, as we decrease the temperature,
parameters monitoring the folding process such as potential energy and root
mean square distance with respect to the native state, $rmsd$, suggest the
presence of intermediates. We determine the kinetic partition temperature
$T_{KP}$ below which we observe two folding intermediates, $I_1$ and $I_2$,
and above which we do not observe accumulation of intermediates.  Below
$T_{KP}$, intermediate $I_1$ forms when the two termini and the strand
following the RT-loop form a $\beta$-sheet, prior to the formation of the
folding nucleus.  This intermediate effectively splits the folding transition
into fast and slow folding pathways.  Dissociation of part of the
$\beta$-sheet leads the protein to the native state. We also find that
stabilization of this $\beta$-sheet and subsequent dissociation of the distal
hairpin may lead to intermediate $I_2$.

The key structural characteristics of intermediate $I_1$ allow us to define a
similarity score function that probes the presence of the intermediate in a
folding transition. We find that $I_1$ is populated even at $T_{KP}$.  This
result suggests that one can obtain information regarding the existence of
putative intermediates by studying the folding trajectories at $T_{KP}$.
However, at this temperature, no intermediates are noticeable if one limits
the analysis only to the distribution of folding times.

We observe that the folding pathways of the model SH3 domain are highly
sensitive to temperature, suggesting the important role of the environmental
conditions in determining the folding mechanism. Our findings suggest that
the SH3 domain, a two-state folder, may exhibit stable intermediates under
extreme experimental conditions, such as very low temperatures.

\section{MATERIALS AND METHODS}

\subsection{Model Protein and Interactions}

We adopt a coarse-grained description of the protein by which each amino acid
is reduced to its C$_\beta$ atom (C$_\alpha$ in case of Gly). Details of the
model, the surrounding heat bath, and the selection of structural parameters
are discussed in detail in a previous study\cite{Borreguero02}.  The
selection of the set of interaction parameters among amino acids is of
crucial importance for the resulting folding kinetics of the model protein
\cite{Plotkin02,Thirumalai02,Pande00}. Experimental and theoretical studies
of globular proteins
\cite{Borreguero02,Ding02,Plaxco98,Munoz99,Fersht00,Alm99,Galzitskaya99,Shea00,Riddle99,Du99,Micheletti99,Clementi00}
suggest that native topology is the principal determinant of the folding
mechanism. Thus, we employ a variant of the G\={o} model of
interactions\cite{Jang02,Thirumalai02,Cieplak01,Shimada01,Hoang00,Zhou99} ---
a model based solely on the native topology --- in which we prevent formation
of non-native interactions, since we are solely interested in the role that
native topology and native interactions may have in the formation of
intermediates. We perform simulations and monitor the time evolution of the
protein and the heat bath with the discrete molecular dynamics algorithm
\cite{Alder59,Zhou97,Allen87,Rapaport97,Grosberg97,Dokholyan98,Dokholyan00}.
The higher performance of this algorithm over conventional molecular dynamics
allows one to increase the computational speed up to three orders of
magnitude.

\subsection{Frequencies and Folding Simulations} 

To calculate the frequency map at $T=1.0$, we probe the presence of the
native contacts in each of the $1100$ initially unfolded conformations. Then,
we compute the probability of each native contact to be present. To calculate
the frequency map at $T_{\rm target}$, we select one particular folding
transition and we probe the presence of the native contacts during the time
interval that spans after the initial relaxation and before the simulation
reaches the folding time $t_F$. To compute $t_F$, we stop the folding
simulation when $90\%$ of the native contacts form. Then, we trace back the
folding trajectory and record $t_F$ when the root mean square distance with
respect to the native state, $rmsd$, becomes smaller than $3$\mbox{\AA}.  We
consider all protein conformations occurring for $t > t_F$ as belonging to
the folded state and of no relevance to the folding transition.

\subsection{Similarity Score Function}

\begin{figure}[htb]

  \centerline{
    \epsfig{file=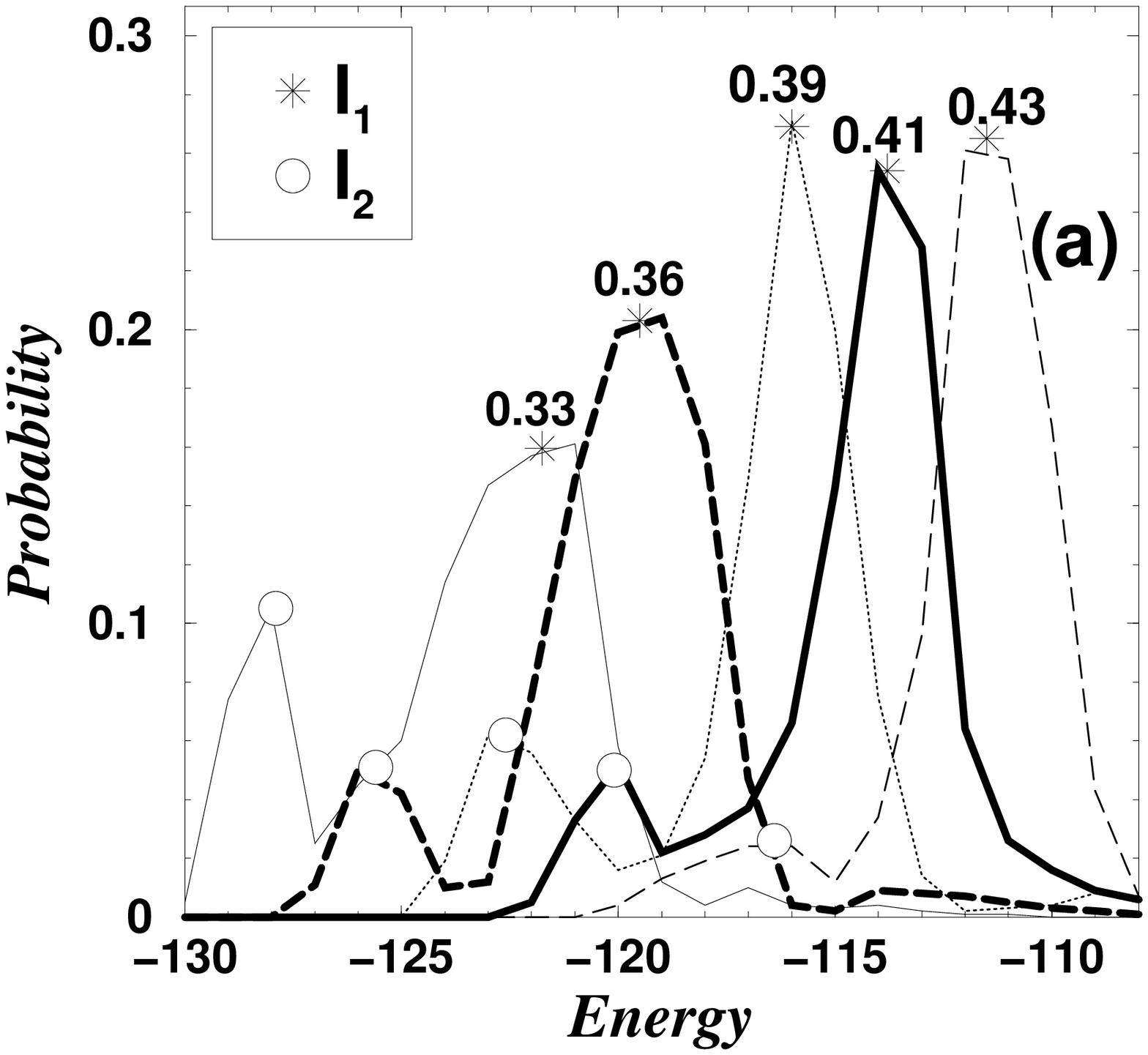, height=79mm, width = 80mm}
    \epsfig{file=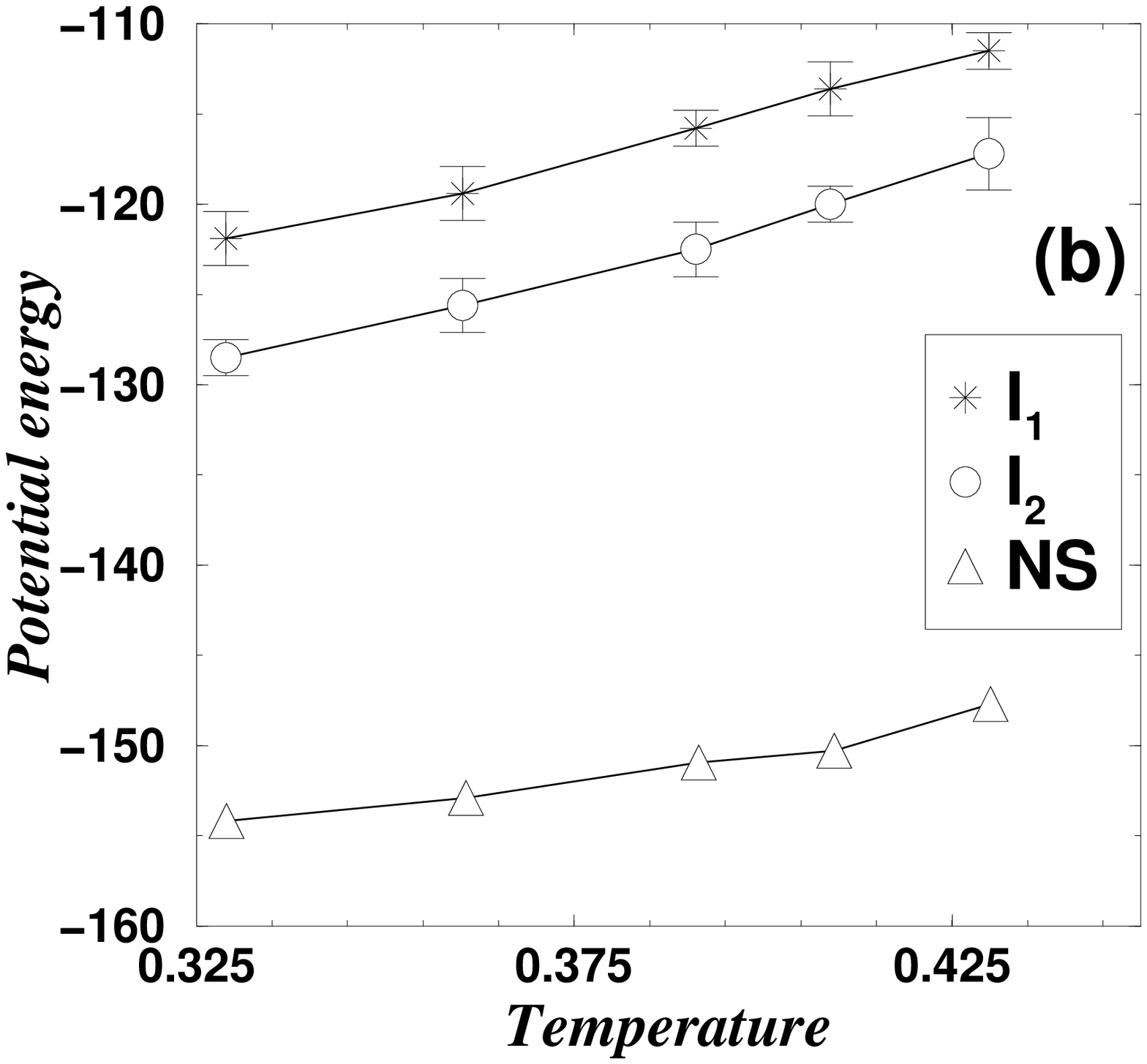,  height=79mm, width = 80mm }
  }

  \centerline{
    \epsfig{file=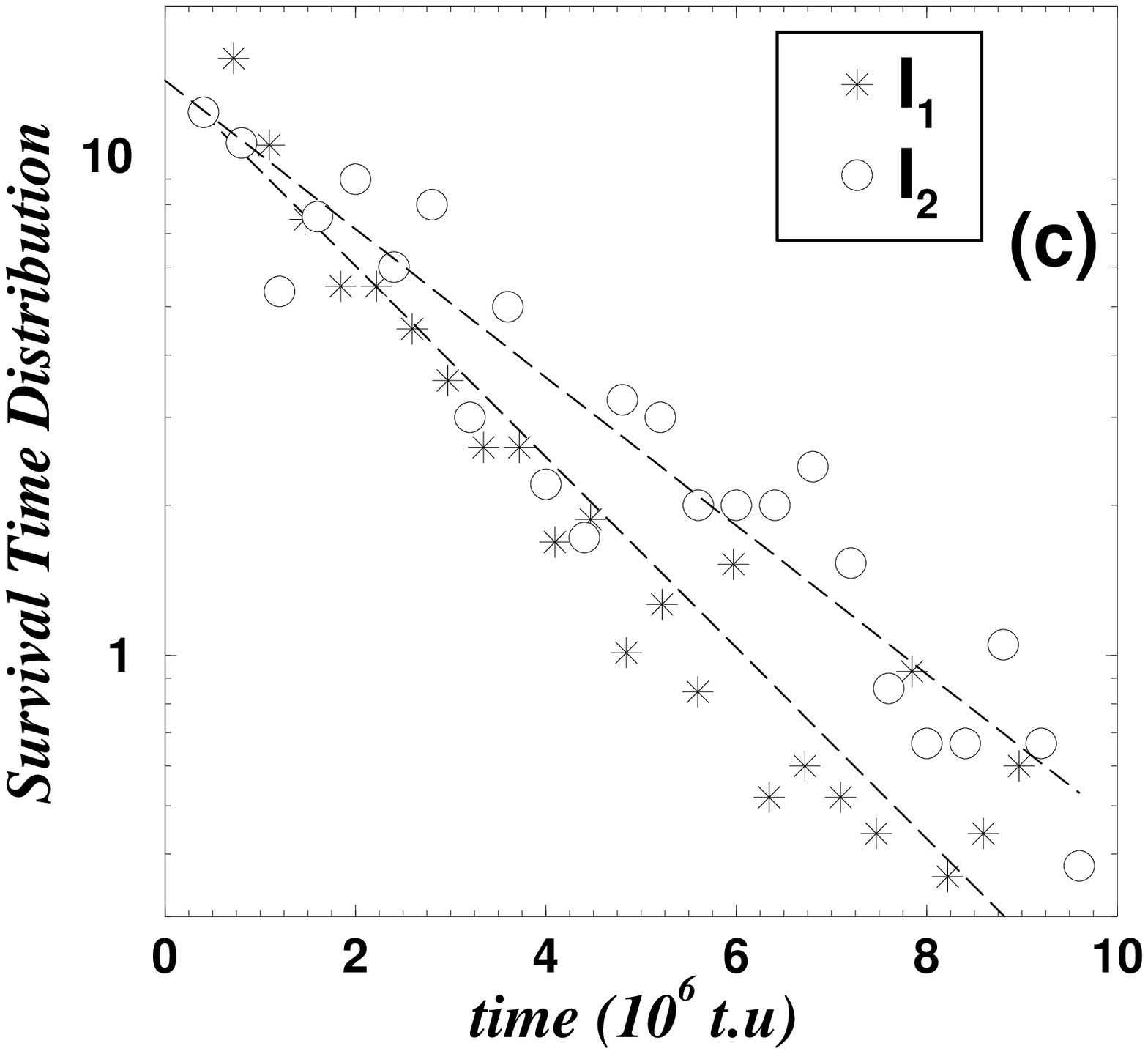,  height=79mm, width = 80mm }
    \epsfig{file=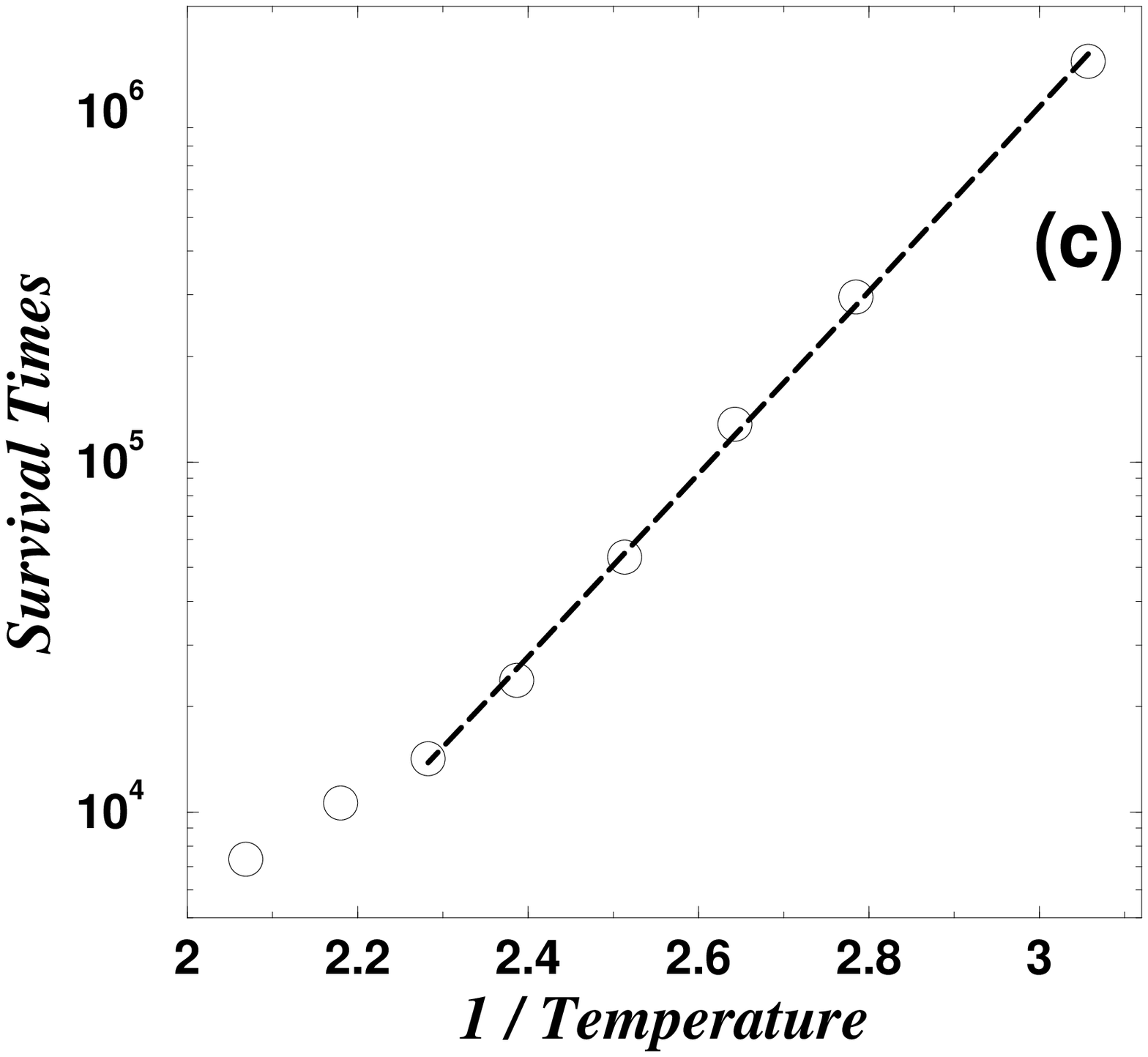,  height=79mm, width = 80mm }
  }

  \centerline{
    \epsfig{file=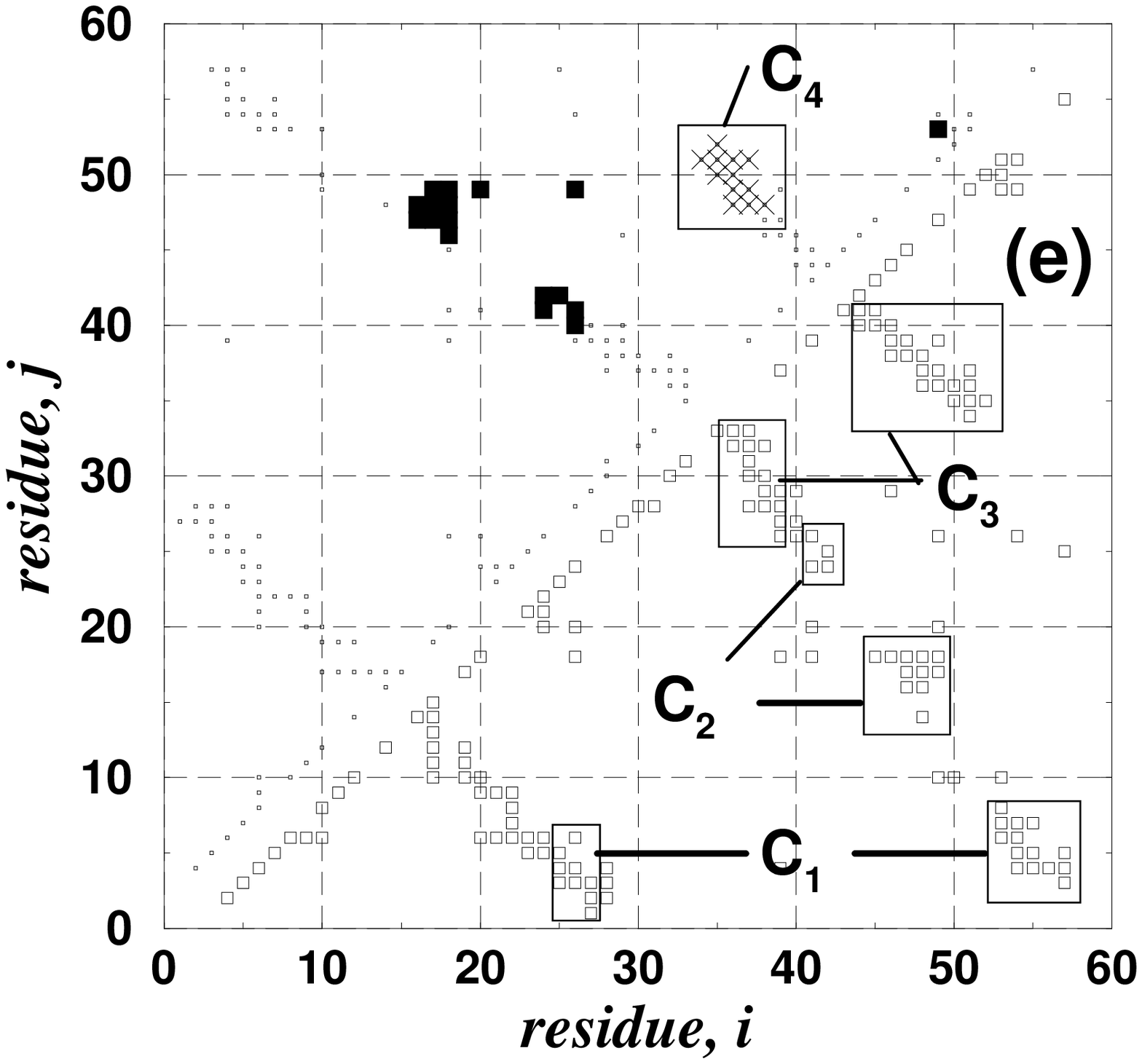,  height=80mm, width = 80mm }
    \epsfig{file=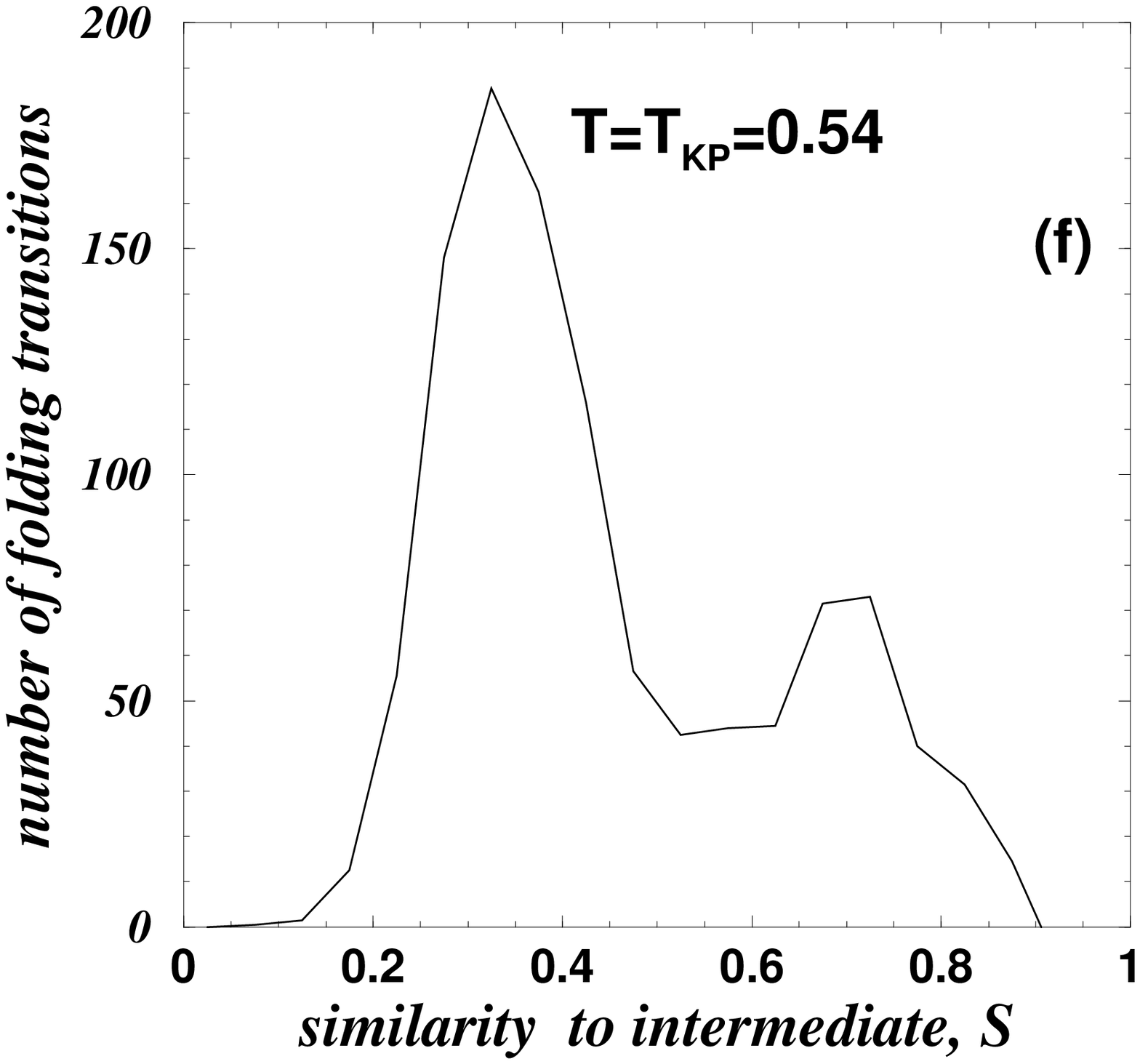,  height=80mm, width = 80mm }
  }

\caption[]{
 }
\label{fig:fig3}
\end{figure}

\begin{figure}[htb]
        
  \centerline{
    \epsfig{file=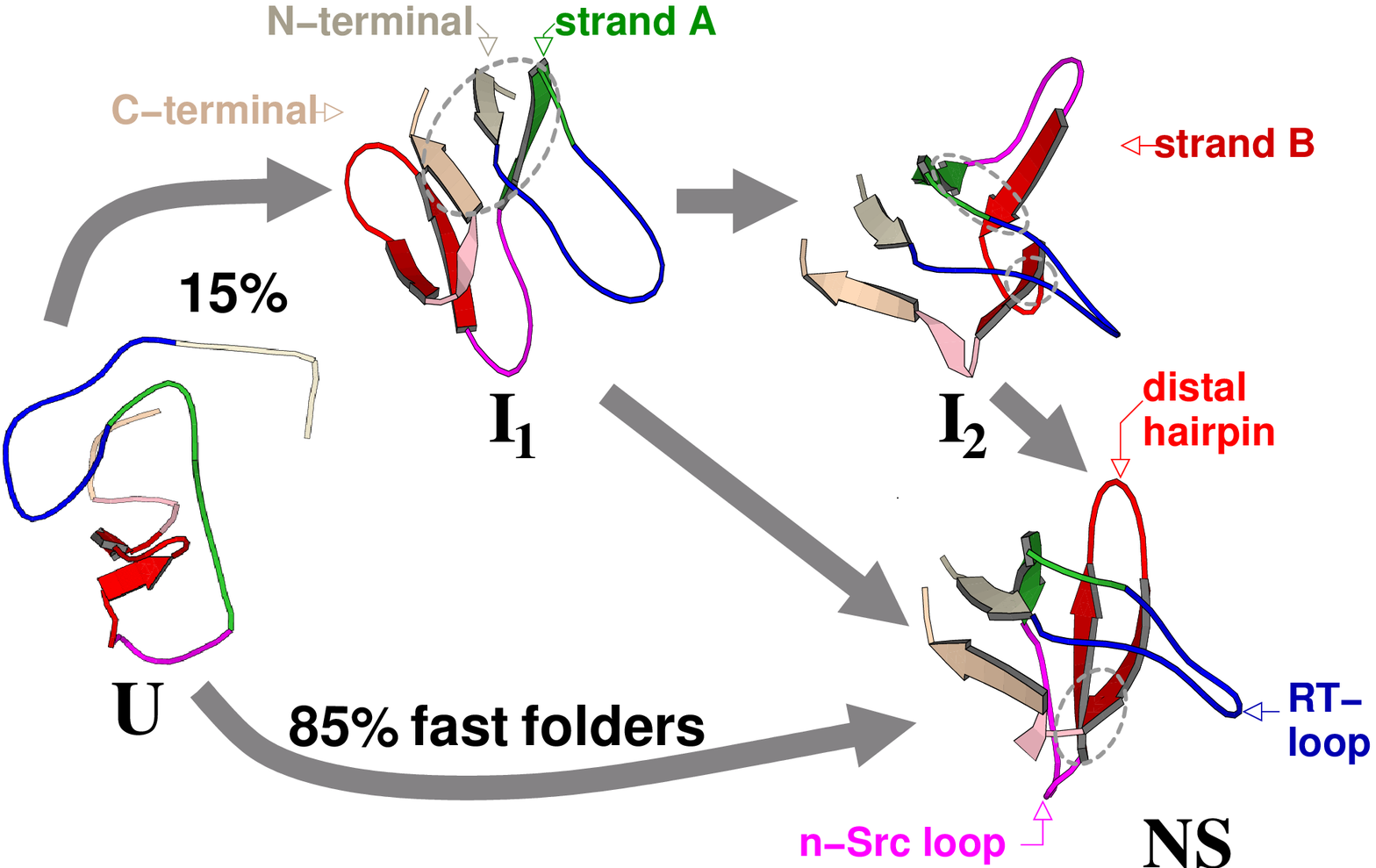, width = 180mm }
  }

\caption[]{
}
\label{fig:pathCartoon}
\end{figure}

We introduce the similarity score function, $S=(a/23)(15-b)/15$, where
$a$ is the number of native contacts belonging to the set of contacts C$_1$,
and $b$ is the number of native contacts belonging to set C$_2$
(Fig.~\ref{fig:fig3}e). C$_1$ has $23$ contacts and C$_2$ has $15$ contacts.
If the protein is unfolded, then $a \approx b \approx 0$, thus $S \approx 0$.
Similarly, if the protein is folded, then $a \approx 23$ and $b \approx 15$,
thus $S \approx 0$ again. Finally, if the protein adopts the intermediate
$I_1$ structure, then $a \approx 23$ and $b \approx 0$, thus $S \approx 1$.

\vspace{10mm} 

We acknowledge E.~I.~Shakhnovich for insightful discussions, R.~Badzey for
careful reading of the manuscript, and the Petroleum Research Fund for
support.

\onecolumn
\newpage
\twocolumn

%bibliography must be located before figures
\bibliographystyle{jmb}
\bibliography{111MYBIB}

\newpage

Figure 1: (a) Upper triangle: c--Crk SH3 domain contact map with $160$ native
contacts. The secondary structure elements are the clusters of contacts that
are organized perpendicularly to the map diagonal.  Long-range contacts
between the two termini are enclosed in the circle, and long-range contacts
between the RT-loop and the distal hairpin are enclosed in the square. Lower
triangle: the frequency map for the initial set of 1100 unfolded
conformations at $T=1.0$.  (b) Frequency map of the unfolded state at
$T=T_F=0.626$.  We compute the frequencies for a particular folding
transition, whose potential energy trajectory we show in the inset (see
Methods). Same for (c) $T=0.54$ and (d) $T=0.33$, the lowest temperature
studied.\\

Figure 2: (a--e) Histograms of folding times for selected temperatures.  At
$T=0.33$ and $T=0.36$, the two lowest temperatures studied, histograms have a
maximum for long folding times ($\downarrow^{\hexstar}$), which suggests the
existence of putative intermediates. At $T=0.33$, a maximum in the histogram
($\downarrow^{+}$), not present at $T=0.36$, corresponds to short lived
kinetic traps. The distributions of folding times are unimodal at higher
temperatures. At $T=0.54$, the histogram is compact, and has no tail of long
folding times. At $T=0.64$, the histogram fits a single-exponential
distribution $e^{-t_F/\langle t_F \rangle}/\langle t_F \rangle$ for times
larger than the relaxation time of $1500$ t.u. (dashed line).  We estimate
the errors of the histogram bars as the square root of each bar.  (f) Average
folding time versus temperature. Each dot represents the folding time for a
particular folding transition.  (g) Ratio $r$ of the average and the standard
deviation, $r=\langle t_F \rangle/\sigma_F$, for the distribution of folding
times. The ratio approaches one above $T_{KP}$ and zero below $T_{KP}$. The
ratio is maximal at $T_{KP}$, indicating a compact
distribution of folding times at this temperature.\\

Figure 3: (a) Distributions of the potential energies of the slow folding
pathways for temperatures below $T=0.43$. The distributions are bimodal,
suggesting two putative intermediates $I_1$ and $I_2$.  (b) The potential
energy of the distribution peaks (\ding{83} and $\ocircle$) increases with
temperature, but the energy difference between peaks remains constant. The
energy of the peaks is significantly larger than the equilibrium energy of
the folded state ($\bigtriangleup$). (c) Distributions of survival times at
$T=0.33$ for the high energy intermediate $I_1$ (\ding{83}), $\langle t_F
\rangle=1.81 \times 10^6$ and $\sigma=1.85 \times 10^6$, and the low energy
intermediate $I_2$ ($\ocircle$), $\langle t_F \rangle=2.47 \times 10^6$ and
$\sigma=2.43 \times 10^6$, fit to single-exponential distributions. (d)
Arrhenius-fit of the average survival time of intermediate $I_2$ below
$T=0.44$. This upper bound temperature coincides with the temperature below
which the distribution of the potential energies (Fig.~\ref{fig:fig3}a) of
the slow folding pathways becomes bimodal. (e) Upper triangle: Absent
contacts (filled squares) and present contacts (crosses, ``C$_4$'') in
intermediate $I_1$.  Upon the transition $I_1 \rightarrow I_2$, these
contacts reverse their presence (so that the filled squares are the present
contacts and the crosses are the absent contacts).  There are five more
squares than crosses, which roughly accounts for the difference of six energy
units between the two intermediates. Lower triangle: long-range contacts
``C$_1$'' are present in intermediate $I_1$, and long-range contacts
``C$_2$'' are absent. There are $23$ contacts in ``C$_1$'' and $15$ contacts
in ``C$_2$''. (f) Probability that a folding transition at $T=T_{KP}=0.54$
contains a protein conformation with
similarity $S$ to intermediate $I_1$ (see Methods).\\

Figure 4: Schematic diagram of fast and slow folding pathways. At $T=0.33$,
approximately in $15\%$ of the simulations, the model protein undergoes a
folding transition through the slow folding pathways. We show the protein
structure in $I_1$ and $I_2$ using the secondary structural elements of the
native state, although some of these elements are not formed. In intermediate
$I_1$, both termini and the strand A form a $\beta$-sheet (in the ellipse).
The corresponding set of native contacts is C$_1$. Dissociation of the
$\beta$-sheet leads to rearrangements of the protein conformation and
successful folding to the native state (NS).  However, dissociation of the
distal hairpin (in red) leads to more localized rearrangements that may lead
the protein to intermediate $I_2$. Upon $I_1 \rightarrow I_2$ transition,
contacts of C$_2$ (the two ellipses in $I_2$) form, but contacts of C$_4$
(the ellipse in NS) break.

\onecolumn
\newpage

\end{document}